\documentstyle[pra,aps,twocolumn,epsfig]{revtex}

\begin{document}


\wideabs{

\title{Dynamics of dark solitons in elongated 
Bose-Einstein condensates}

\author{A. E. Muryshev$^{1}$, G. V. Shlyapnikov$^{1,2,3}$,  W. Ertmer$^{4}$
K. Sengstock$^{4,5}$,
and M. Lewenstein$^{6}$}

\address{(1) Russian Research Center Kurchatov Institute, Kurchatov Square, 123182  Moscow, Russia\\  
(2) FOM Institute for Atomic and Molecular Physics,
Kruislaan 407, 1098 SJ Amsterdam, The Netherlands\\
(4) Laboratoire Kastler Brossel$^{*}$, 24 rue Lhomond, F-75231 Paris Cedex 05, France\\
(3) Institut f\"ur Quantenoptik, Universit\"at Hannover,
D-30167 Hannover,
Germany\\
(4) Institut f\"ur Laserphysik, Universit\"at Hamburg, 20355 Hamburg,
Germany\\
(5) Institut f\"ur Theoretische Physik, Universit\"at Hannover,
D-30167 Hannover,
Germany
}

\maketitle

\begin{abstract}

We find two types of moving dark soliton textures in elongated
Bose-Einstein condensates: {\it non-stationary kinks} and {\it proper 
dark solitons}. The former have a curved notch region and rapidly decay 
by emitting  phonons and/or proper dark solitons. The proper moving 
solitons are characterized by a flat notch region and we obtain the 
diagram of their dynamical stability. At finite temperatures the 
dynamically stable solitons decay due to the thermodynamic instability. 
We develop a theory of their dissipative dynamics and explain  
experimental data.

\end{abstract}


}


Recently, several spectacular experiments have demonstrated the creation
of vortices \cite{jila,ens,mit} 
and dark solitons \cite{nist,hannover,anderson} in
Bose-Einstein condensates (BEC) of trapped alkali-atom  gases. 
These experiments open an unprecedented possibility to study 
the dissipative dynamics 
of such macroscopically excited Bose-condensed states. The literature 
concerning the dynamical and thermodynamic stability of vortices is quite 
extensive (see, e.g., \cite{fetter} for review).  
As the vortex 
has a topological charge (circulation), the dynamically stable 
(single-charged) vortices can decay only when reaching the border of the 
condensate. At finite temperatures in non-rotating traps, the motion of 
the vortex core towards the  border is induced by the 
interaction of the vortex with thermal excitations and is rather slow. 
The lifetime of vortices in trapped 
condensates \cite{vortheory2,vortheory3} is thus  relatively long and
extends to a few seconds in the experiments \cite{jila,ens,mit}.

Dark solitons have a density dip and a phase slip in one direction and, 
as well as vortices, they are particular solutions of the 
Gross-Pitaevskii (GP) equation. Extensive studies 
of dark solitons in nonlinear optics \cite{nonlin} have 
expounded their transverse dynamical instability in 3D geometries, 
leading to the  undulation of the soliton plane and decay into 
vortex-antivortex pairs and phonon waves. This scenario is similar to that
observed at NIST \cite{nist} for solitons in almost  spherical trapped
BEC's. Recently, the decay of solitons into vortex rings  was 
observed at JILA \cite{nist}. 
The transverse instability of dark solitons can be suppressed by a strong 
radial confinement of the soliton motion in elongated
traps \cite{muri1}. Dynamically stable
solitons in such traps  are not, however, 
thermodynamically stable, and
their dissipative dynamics is expected to be fundamentally
different from that of vortices.

In contrast to vortices, the soliton has no topological charge and
can decay without reaching the border of the condensate. Dark 
solitons behave as objects with a negative mass. 
The scattering of thermal excitations from the soliton decreases its 
energy, and the soliton accelerates towards the speed of sound, gradually 
looses its contrast, and ultimately disappears. This mechanism has been 
proposed in Ref. \cite{muri2}, and the lifetime of the soliton has been 
obtained in terms of the reflection coefficient of the excitations. 
However, the theory of Ref. \cite{muri2} pertains 
to the 1D case, where the GP equation is integrable and the reflection 
coefficient is  strictly  zero within the Bogolyubov approach. 
Thus, one expects very long lifetimes of solitons in this limit.
On the other hand, 
in 3D elongated traps the GP equation is no longer integrable and the
scattering of thermal excitations from the soliton should be efficient.
The absence of topological charge and integrability should then
lead to a much faster dissipative dynamics of solitons than that of vortices.
The Hannover results \cite{hannover} indeed
suggest that moving solitons generated in a cigar-shaped trap are
dynamically stable, but
their  contrast decreases to zero  within typically
$\simeq 15$ms as a result of the  thermodynamic
instability. 

In  this Letter we study the dynamics of moving dark 
solitons in 3D elongated Bose-Einstein condensates and present three
important results:  i) We find that using the "phase imprinting method"
\cite{hannover,dobrek} one can generate at least two kinds of soliton
textures: {\it non-stationary kinks} and {\it proper dark solitons}. The
former have a notch region  that moves with radially non-uniform velocity
and undergoes bending similar to
that observed at NIST  \cite{nist}. These textures are dynamically unstable
and decay  via the emission of phonons and/or proper dark solitons. The proper 
solitons are characterized by a flat notch region and propagate without 
changing their shape;
ii) We derive the diagram of dynamical stability for proper 
solitons; 
iii) We solve the problem of reflection of  
excitations from the soliton and analyze its decay due to
thermodynamic instability. The dissipative dynamics exhibits an
interplay between the extent of non-integrability and the absence of
topological charge, and the soliton
lifetime ranges from milliseconds for Hannover-type 3D
solitons to more than seconds in quasi1D  geometries.

We  consider a condensate with repulsive 
interaction (the scattering length $a>0$). The condensate 
wave function can be written as $\Psi({\bf r},t)\exp{(-i\mu t)}$, where 
$\mu$ is the chemical potential. In an infinitely long cylindrical harmonic
trap this function satisfies  the GP equation 
\begin{equation}        \label{GPE}
\!\!i\hbar\frac{\partial\Psi({\bf r},t)}{\partial t}\!=\!\! 
\left\{\!-\frac{\hbar^2}{2m}\Delta\!+\!\frac{m}{2}\omega_{\rho}^2\rho^2\!+
\!g|\Psi({\bf r},t)|^2\!\!\!-\!\!\mu\!\right\}\!\!\Psi({\bf r},t).\!\!  
\end{equation}
Here $\omega_{\rho}$ is the frequency of the radial ($\rho$) confinement,
$g=4\pi\hbar^2a/m$, and $m$ is the atom mass. The wave function of the
ground-state condensate minimizes the corresponding energy functional and 
is the solution of Eq.(\ref{GPE}) with zero lhs. Macroscopically excited 
Bose-condensed states (solitons, vortices etc.) are described by any other 
stationary, or time dependent solution of  Eq.(\ref{GPE}). Obviously,
they  are thermodynamically unstable as they do not correspond to the
minimum of  the energy functional. 

A stationary, or solitary-wave macroscopically excited BEC state can
also be dynamically unstable with regard to elementary excitations
around it. The unstable excitation modes are characterized by complex 
eigenfrequencies and grow exponentially in time, which indicates that the
BEC state will evolve far from the initial shape. 

Strictly speaking, dark solitons are solutions of the 1D GP equation 
in free space. They are characterized by a local density minimum (notch)
moving with a constant velocity $v$, and by a phase gradient of the 
wave function at the position of the minimum.  
 In an otherwise uniform condensate of 
density $n_0$, the dark soliton state is described by the wave function 
(see \cite{muri2} and refs. therein)
\begin{equation} 
\Psi(z,t) = \sqrt{n_0}\left(\cos{\theta}-i
\sin{\theta} 
\tanh\left[\sin{\theta}(z-vt)/l_0\right]\right) ,
\label{soliton_eqn}
\end{equation}
where $\cos{\theta}={v}/{c_s}$, and $c_s=\sqrt{n_0g/m}$ is the speed 
of sound.
The quantities $\theta$ and $-\theta$ are the phases of the soliton 
state at
$(z-vt)\rightarrow -\infty$ and $(z-vt)\rightarrow\infty$, 
respectively. The width $L$ of the notch (soliton plane) is of the order of
the correlation length 
$l_0=\hbar/mc_s$. Approximate soliton solutions can be also found in 
1D harmonic traps \cite{dum,busch}. Solitons which have non-zero velocity in
the  center of the trap oscillate along the trap axis \cite{busch}. 

In 3D harmonic traps the solutions of the GP equation, describing standing
dark solitons ($v=0$ and $\partial\Psi/\partial t=0$), have been found in 
\cite{muri1,feder}. For infinitely long cylindrical condensates these solutions
follow from Eq.(\ref{soliton_eqn}) and can decay due to the transverse
dynamical instability. The stability criterion requires a strong radial 
confinement providing a non-Thomas-Fermi (TF) regime with the radial size of
the condensate $r\alt L\sim l_0$. 

The existence of moving soliton-like textures in 3D elongated 
condensates is
confirmed by the experiments and simulations \cite{hannover}, but no
analytical solution has been found so far. In the TF regime
($\mu\gg\hbar\omega_{\rho}$), one
can use Eq.(\ref{soliton_eqn}) with $\rho$-dependent $n_0$ and $l_0$.
Then, the absence of the radial flux of particles at an infinite
axial separation from the notch requires the phase $\theta$ to be independent
of the radial coordinate. This means that the notch velocity $v$ depends on
$\rho$ and is proportional to the local velocity of sound $c_s(\rho)$. Hence,
the central regions of the notch move faster than the borders, and an initially
flat notch region starts bending in the course of motion. Our simulations for
TF condensates show that such {\it non-stationary kinks} can be created by
phase imprinting with $\rho$-independent optical potential. 
They are dynamically unstable and decay on a time scale of the order of
$\omega_{\rho}^{-1}$. The notch surface bends more and more, whereas the
notch velocity increases and the depth decreases. Ultimately, the 
non-stationary kink decays emitting excitation waves and 
(for $\mu/\hbar\omega_{\rho}\alt 10$) a {\it proper dark
soliton}: The latter moves with $\rho$-independent velocity and does not
change its shape characterized by a flat notch region.  
In non-TF condensates ($\mu\!\sim\!\hbar\omega_{\!\rho}$) we generated the
proper solitons directly  by simulating the phase imprinting. 

In the absence of dissipation, the velocity $v$ of a proper soliton 
remains constant in an
infinitely long cylindrical condensate. The wave function $\Psi$
depends on $\rho$  and  $x=z-vt$. In order to find this wave function, we
write it in the form $\Psi(\rho,x) =\psi(\rho,x)f(x)$, where the functions
$\psi(\rho,x)$ and $f(x)$ satisfy the equations  
\begin{eqnarray}    
\!i\hbar\partial\psi/\partial t&=&\left\{\!-(\hbar^2/2m)(\Delta\!
+2(\nabla_xf/f)\nabla_x)\!\!+
m\omega_{\rho}^2\rho^2/2\!\right. \nonumber\\ &+&\!g|f|^2|\psi|^2\!
\!-\left.\!\tilde\mu(|f|)
\right\}\psi,\!   \label{eqrho} \\
i\hbar\partial f/\partial t &=& -(\hbar^2/2m)
\Delta_x f
+(\tilde\mu(f)-\mu)f,  \label{eqf}
\end{eqnarray}
The quantity $\tilde\mu$ is a functional of $f$ and has to be found 
self-consistently from Eqs.~(\ref{eqrho}) and (\ref{eqf}). We will select
the function $\psi(\rho,x)$ such that at infinite $x$ it becomes the wave
function of the ground-state condensate, $\psi_0(\rho)$. Hence, for
$|x|\rightarrow\infty$ we have $|f|\rightarrow 1$ and
$\tilde\mu\rightarrow\mu$. 

We consider the limiting case where the axial size $L$ of the soliton
notch greatly exceeds the radial size $r$ of the condensate. Then the
radial distribution of particles is close to $n_0(\rho)=\psi_0^2(\rho)$ for the
ground-state condensate. The quantity $\tilde\mu(f)$ is close to
$\mu$ and can be expressed as  
$\tilde\mu(f)=\mu+(|f|^2-1)g\partial\mu/\partial
g+\delta\tilde\mu$, where $\delta\tilde\mu$ is a 
correction of higher order in $r/L$.
The quantity $g\partial\mu/\partial g=m{\bar c}_s^2$, where ${\bar c}_s$ is 
nothing else than the velocity of axially propagating sound waves in the 
ground-state condensate. In the quasi 1D regime, where the interparticle
interaction at maximum condensate density
$n_{0m}g\ll\hbar\omega_{\rho}$, we have an almost Gaussian  density profile 
$n_0(\rho)=n_{0m}\exp{(-\rho^2/l_{\rho}^2)}$. The radial size $r\sim
l_{\rho}=(\hbar/m\omega_{\rho})^{1/2}$, and the small parameter of 
the expansion for $\tilde\mu$ is $(r/L)^2\sim
n_{0m}g/\hbar\omega_{\rho}$. 
To first order in $n_{0m}g/\hbar\omega_{\rho}$ we obtain
$g\partial\mu/\partial g=n_{0m}g/2$. This gives the velocity ${\bar c}_s$ 
which is by a factor of $\sqrt{2}$ smaller than the speed of sound at maximum 
density: ${\bar c}_s=\sqrt{n_{0m}g/2m}$. 
For radially TF condensates the chemical potential $\mu\propto\sqrt{g}$ and
we arrive at the same expression for ${\bar c}_s$. 
This result for TF elongated condensates has been obtained in 
\cite{axsound} and found in the MIT experiment \cite{mitsound}. 
The condition $r\ll L$ requires fast TF solitons for which the density 
dip is small and the function $|f(x)|$ is close to 1.

Omitting the higher order correction $\delta\tilde\mu$, Eq.(\ref{eqf}) for 
the function $f(x)$ becomes an ordinary 1D GP equation
\begin{equation}     \label{eqff0}
i\hbar\partial f/\partial t=-(\hbar^2/2m)\Delta_x f
+m{\bar c}_s^2[(|f|^2-1)f.
\end{equation}
The terms in the rhs of Eq.(\ref{eqff0}) are related to
the axial kinetic energy and to the mean-field interparticle interaction in the
presence of the radially inhomogeneous density profile of the Bose-condensed
state. The dark soliton   
solution $f(x)$ is then given by the rhs of Eq.(\ref{soliton_eqn}), where  
$c_s$ is replaced by ${\bar c}_s$, $l_0$ by ${\bar l}_0=l_0/\sqrt{2}$, and 
$n_0$  by unity. The soliton-like 
wave function of the condensate can be written as 
$\Psi(\rho,x)=\psi_0(\rho)f(x)$. This clearly shows that the maximum soliton 
velocity (at which the soliton disappears) is equal to ${\bar c}_s$, i.e. is
by  $\sqrt{2}$ smaller than the  sound velocity at maximum condensate density. 

In the  quasi1D regime, to first order in $n_{0m}g/\hbar\omega_{\rho}$ the 
function $\psi(\rho,x)=\psi_0(\rho)+\delta\psi(\rho,x)$, where a small
term $\delta\psi$ is real. The second order correction 
$\delta\tilde\mu$ is then equal to 
$\frac{1}{2}(|f|^2-1)]^2g^2\partial^2\mu/\partial g^2$. 
We find $g^2\partial^2 \mu/\partial g^2=-\gamma
n_{0m}g$, with $\gamma\!=\!3n_{0m}g\ln(4/3)/4\hbar\omega_{\rho}\!\ll\!1$. 
Thus, we obtain 
$\tilde\mu(f)\!=\!\mu\!+\!m{\bar c}_s^2[(|f|^2\!-\!1\!)\!-
\!\gamma(|f|^2\!-\!1)^2]$, and 
Eq.(\ref{eqf}) becomes
\begin{equation}     \label{eqff}
i\hbar\frac{\partial f}{\partial t}=-\frac{\hbar^2}{2m}\frac{d^2f}{dx^2}+
m{\bar c}_s^2[(|f|^2-1)-\gamma(|f|^2-1)^2]f.
\end{equation}  

The quasi1D dark solitons are dynamically stable, and we now discuss their 
thermodynamic instability in the presence of a thermal cloud. 
With $\gamma=0$, Eq.(\ref{eqff}) 
transforms to Eq.(\ref{eqff0}) which is integrable. Hence, the dark solitons 
described by this equation are transparent for 
thermal excitations. There is no energy and momentum exchange between the
soliton and the thermal cloud, and the thermodynamic instability does not
manifest itself. The dissipative dynamics of the solitons originates from the
interaction between the radial and axial degrees of freedom, which 
in the quasi 1D regime is described
by the small term $\gamma(|f|^2-1)^2$ in the rhs of Eq.(\ref{eqff}). This term
only slightly modifies the wave function $\Psi(\rho,x)$, but it lifts the 
integrability of the equation and is responsible for the reflection
of excitations from the soliton.

For finding the reflection coefficient we have solved the Bogolyubov-de Gennes
equations following from Eq.(\ref{eqff}). 
For the phonon branch of the spectrum, 
where the axial momentum of an excitation $k\ll {\bar l}_0^{-1}$, the
excitation wave functions $u,v$ were found in the form of expansion in powers
of $k{\bar l}_0$ and $\gamma$ around the fundamental modes of the Bogolyubov-de
Gennes equations with $\gamma=0$. The $u,v$ functions of these equations were
obtained straightforwardly for an arbitrary $k$ and used for
calculating the reflection coefficient $R(k)$ from the Fermi golden rule. For
$k{\bar l}_0\ll 1$ the obtained $R(k)$ matches the one following from the 
method of fundamental modes. At any energy $\varepsilon$ 
and momentum $k$ of the incident wave we have 
\begin{equation}          \label{R}
\!\!\!R(k)\!\!=\!\!\left[\!\frac{8\pi\gamma(\varepsilon\!-\!\hbar kv){\bar
l}_0^2}{9\hbar\sinh\{(\pi(|k|\!\!+\!\!|k'|){\bar
l}_0/2\sqrt{1\!\!-\!\!v^2\!/\!{\bar
c}_s^2}\}}\!\right]^{\!2}\!\!\!\!\!\frac{|kk'|}{\nu(k)\nu(k'\!)},\!\! 
\end{equation}    
where $\nu(k)=\partial(\varepsilon-\hbar kv)\partial\hbar k$
is the group velocity. The energy $\varepsilon'$ and momentum $k'$ of the
reflected wave are related to $\varepsilon$ and $k$ by the energy
conservation law in the reference frame moving together with the soliton, 
$\varepsilon-\hbar kv=\varepsilon'-\hbar k'v$. For small $k$
the reflection coefficient increases as $k^2\propto\varepsilon^2$. The
coefficient reaches its maximum at $k\sim {\bar l}_0^{-1}\sqrt{1-v^2/{\bar
c}_s^2}$ and decays exponentially for large $k$. The calculations leading to 
Eq.(\ref{R}) are quite lengthy and will be published elsewhere.   

The reflection of excitations from the soliton provides a momentum transfer
from the thermal cloud to the soliton. Hence, there is a friction force acting 
on the soliton. The momentum transfer per unit time is given by 
\begin{equation}      \label{mtr}
\dot p=\int_{-\infty}^{\infty}\hbar(k-k')R(k)\nu(k)N(\varepsilon-\hbar
kv)dk/2\pi,
\end{equation}
where $N(\varepsilon-\hbar kv)$ are equilibrium occupation numbers for the
excitations.  

The energy of the soliton can be written in the form $H\!=\!(M{\bar
c}_s^2/3)(1\!-\!v^2/{\bar c}_s^2)^{3/2}$, with $M\!=\!2\pi n_{0m}{\bar l}_0r^2m$
being the effective mass of the soliton ($r\!=\!\sqrt{2}l_{\rho}$ in the 
quasi1D regime).
The soliton energy decreases with increasing $v$. For example, for $v\ll {\bar
c}_s$ we have $H=M{\bar c}_s^2/3-Mv^2/2$. The quantity $N_*=M/m=2\pi
n_{0m}{\bar l}_0r^2\gg 1$ is the number of particles that one has to
remove from the condensate in order to create the soliton density dip. Thus,
the dark soliton can be treated as a heavy classical particle-like object with
a negative mass, and the friction force accelerates
the soliton towards the velocity of sound (see \cite{muri2}). The Hamiltonian
equation $\partial H/\partial p=v$ gives $\dot p=-M\dot v(1-v^2/{\bar
c}_s^2)^{1/2}$, and using Eq.(\ref{mtr}) we obtain the time dependence of the
soliton velocity.    

For $T\gg n_{0m}g$ the time $t$ at which the soliton contrast $C=(1-v^2/{\bar
c}_s^2)$ decreases from the initial value $C_0$ to $C(t)$, can be found from
the relations 
\begin{equation}        \label{tau} 
F(C)-F(C_0)=t/\tau,\,\,\,\,\tau=\hbar N_*/TR_0,
\end{equation} 
where $F(C)$ is a universal function of the contrast, and $R_0=0.084\gamma^2$ 
is the maximum value of the reflection coefficient in the limit of 
$v\rightarrow 0$. The function $F(C)$ was calculated numerically and can be 
approximated as $F(C)\approx 0.47\ln\{(1-C)/C\}$ \cite{zerov}. The quantity
$\tau$ can be regarded as a characteristic lifetime of the soliton. For
example, if the contrast is initially equal to $30\%$, it decreases to $10\%$
at a time $t\approx 0.25\tau$.     

The dissipative dynamics of quasi1D solitons is governed by the small extent of
non-integrability of Eq.(\ref{eqff}) ($\gamma\ll 1$) and the time $\tau$ can 
be very long. 
The physical picture changes to the opposite one if the axial size $L$ of the
notch   becomes comparable with the radial size $r$ of the condensate. Then the
non-integrability of the GP equation is essential and the absence of topological 
charge provides a fast dissipative dynamics. In this case, which corresponds 
to the border of dynamical stability of the soliton, the developed analytical 
approach can be no longer used and we have found the proper soliton
solutions numerically.  
To investigate the dynamical stability of moving proper solitons we have
numerically solved a time-dependent equation for elementary excitations
around the obtained soliton wave function. For a given soliton velocity $v$, 
the axial size of the notch decreases with increasing the ratio 
$n_{0m}g/\hbar\omega_{\rho}$.
Above a critical value $n_{0m}g/\hbar\omega_{\rho}=\xi_c$ the transverse
instability was manifesting  itself in our calculations as a dramatic rise of
excitation modes. 
In Fig.1 we present the critical
ratio $\xi_c$ as a function of the soliton  velocity. For
$n_{0m}g/\hbar\omega_{\rho}<\xi_c$ the solitons are dynamically stable. Note
that $\xi_c=2.5$ for a standing  soliton, which agrees with the earlier
calculation  \cite{muri1}. For the solitons with high velocities  the
stability condition is more relaxed. 

For large $n_{0m}g/\hbar\omega_{\rho}$, i.e. in the TF limit, the soliton
velocity $v$ required for the transverse dynamical stability approaches 
the sound velocity ${\bar c}_s$.
If $n_{0m}g/\hbar\omega_{\rho}>10$, then the transverse instability is
suppressed only for $v$ larger than the Landau critical velocity $v_*$
calculated for elongated condensates in \cite{fed}. Solitons with $v>v_*$ are
characterized by the longitudinal instability related to the Cherenkov
radiation of axially propagating excitations. We have found this 
instability numerically and established that it develops on a time scale
longer than that of the transverse instability. Thus, for $n_{0m}g
/\hbar\omega_{\rho}>10$ dark solitons are always dynamically unstable.

\vspace{0.4cm}
\begin{figure}
\begin{center}
   \epsfxsize 4.5cm
   \epsfbox{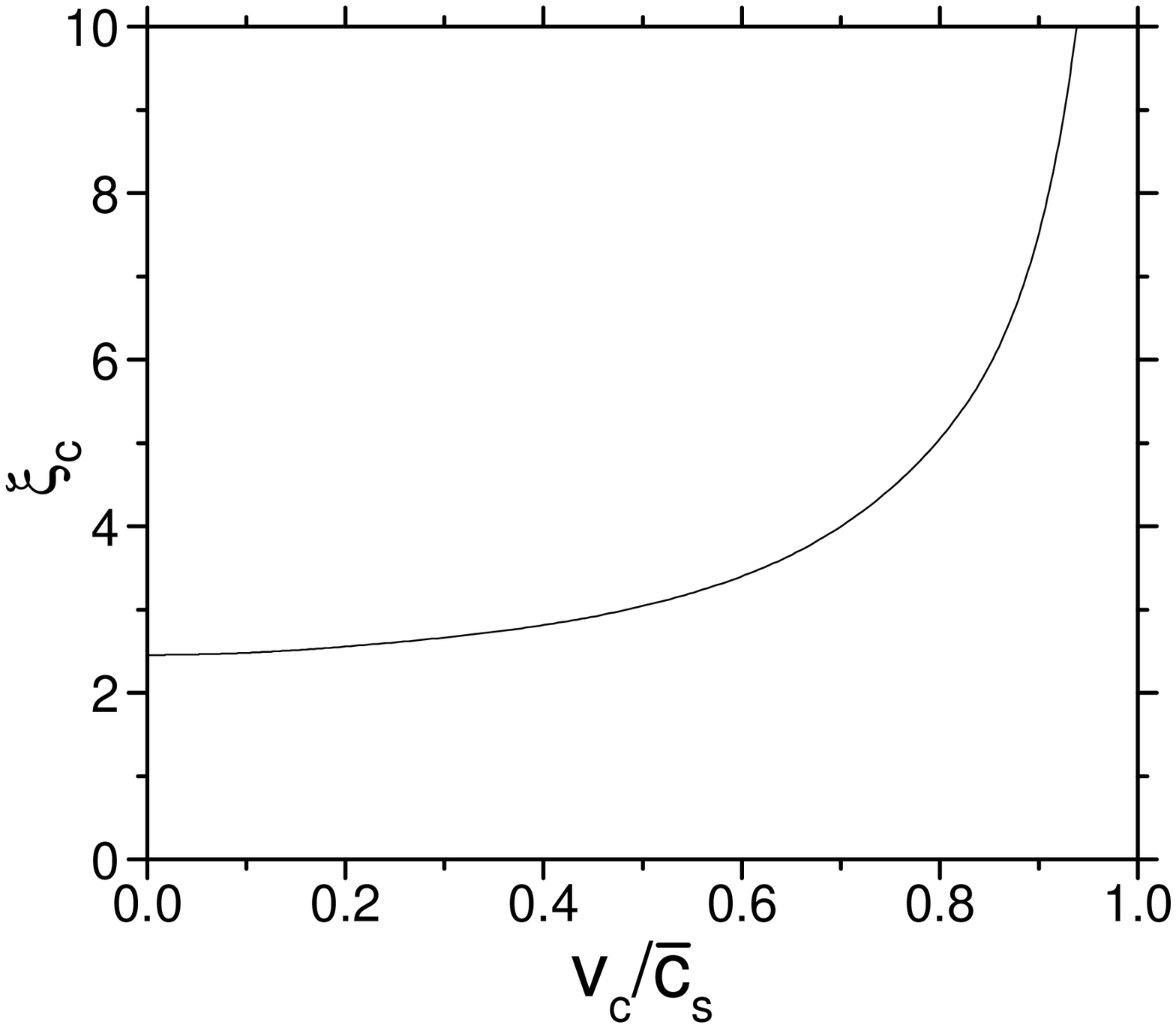}   
\end{center}
\caption {\small 
The critical ratio $n_{0m}g/\hbar\omega_{\rho}=\xi_c$ versus the soliton 
velocity $v$ (in units of ${\bar c}_s$). Solitons are stable 
below this curve.}  
\label{stab}
\end{figure}

The condition $L\sim r$ was fulfilled for solitons in the Hannover 
experiment \cite{hannover}. 
For the number of atoms \( N\approx
1.5\times 10^{5} \), and the trap frequencies \( \omega _{z}=2\pi \times 14
\)Hz  and \( \omega _{\rho }=2\pi \times 425 \)Hz, we calculate the critical 
temperature  \( T_{c}\approx 350 \)nK, the
maximum density $n_{0m}\approx 4\times 10^{14}$ cm$^{-3}$, and the chemical
potential $\mu\approx 140$ nK. This indicates that the solitons were in
the TF regime, with $\mu/\hbar\omega_{\rho}\approx 7$. The thermal fraction was
about $10\%$, which corresponds to $T\approx 0.5T_{c}$ and 
$\mu /T\approx 0.8$. The soliton contrast was decreasing from
approximately $30\%$ to below the resolution limit ($10\%$) at a time of 15
ms. The results in Fig.1 indicate that the dark solitons of Ref. 
\cite{hannover} were dynamically stable. By using the  numerical 
technique which will be described elsewhere, we calculated the reflection 
coefficient of excitations from the soliton.
The dependence of $R$ on $k$ and $v$ is similar to that in the limit of $r\!\ll\!L$.
The maximum reflection coefficient for $v\!\rightarrow\!0$ is $R_0\!\approx\!
0.7$. Then from Eq.(\ref{tau}) we find $\tau\approx 80$ ms and conclude
that the soliton contrast decreases from $30\%$ to $10\%$  at a time of 20 ms,
in agreement with \cite{hannover}.

In conclusion, we have investigated the dynamical stability and dissipative
dynamics of solitons in elongated BEC's, and explained the experimental data 
of Ref. \cite{hannover}. 
For recently achieved quasi1D BEC's,
\cite{1D} the parameter $\gamma\sim 0.1$ 
and our theory predicts the soliton lifetime larger than seconds. 
This opens prospects for studying dissipative phenomena originating from the 
quantum character of the boson field omitted in the common GP approach.
Temperature dependence of 
the soliton lifetime offers interesting 
possibilities of BEC thermometry. 

We acknowledge support from Deutsche Forschungsgemeinschaft, 
the Dutch Foundations NWO and FOM, Alexander von Humboldt Stiftung, the
Russian Foundation for Fundamental Research. 
We thank L. Carr, J. Dalibard, P. Fedichev, 
B. Phillips, and A. Sanpera
for discussions.

$^{*}$LKB is an unit\'{e} de recherche de l'Ecole Normale 
Sup\'{e}rieure et de l'Universit\'{e} Pierre et Marie Curie, associ\'{e}e 
au CNRS.


\begin{references}
\bibitem{zerov} This expression is not valid for $v\simeq 0$, where 
the soliton motion starts in the diffusive regime, see \cite{muri2}.
\end{references}
\end{document}